\documentclass[9pt,twocolumn,twoside]{opticajnl}
\journal{opticajournal} 

\setboolean{shortarticle}{true}

\newcommand{\Ea}{\ensuremath{{\cal E}_1}}

\newcommand{\Ec}{\ensuremath{{\cal E}_3}}

\newcommand{\Ed}{\ensuremath{{\cal E}_{1,2}}}
\newcommand{\Er}{\ensuremath{{\cal E}_{\rm R}}}

\newcommand{\Eoo}{\ensuremath{{\cal E}_{2,3}}}

\usepackage{graphicx}
\usepackage{dcolumn}
\usepackage{bm}

\usepackage{lineno}
\usepackage{float}

\title{Exciton coherence propagation measured with non-local four-wave mixing micro-spectroscopy}

\author[1]{M.~Raczyński}
\author[1]{A.~Dydniański}
\author[1]{K.~E.~Połczyńska}
\author[1]{G.~Szwed}
\author[1]{A.~Szczerba}
\author[2,3]{J.-W.~Jung}
\author[2]{G.~Nogues}
\author[4]{W.~Langbein}
\author[1]{P.~Kossacki}
\author[1]{W.~Pacuski}
\author[2,1,5,*]{J.~Kasprzak}

\affil[1]{Institute of Experimental Physics, Faculty of Physics, University of Warsaw, ul. Pasteura 5, 02-093 Warsaw, Poland}
\affil[2]{Université Grenoble Alpes, CNRS, Grenoble INP, Institut Néel, 38000 Grenoble, France}
\affil[3]{Department of Opto-Mechatronics Engineering and Cogno-Mechatronics Engineering, Physics Education,
RCDAMP, Pusan Nat’l University, Busan 609-735, Republic of Korea}
\affil[4]{School of Physics and Astronomy, Cardiff University, Cardiff, CF24 3AA, UK}
\affil[5]{Japanese-French lAboratory for Semiconductor physics and Technology (J-FAST), CNRS–Université Grenoble Alpes–Grenoble INP–University of Tsukuba, 1-1-1 Tennoudai, Tsukuba, 305-8573, Japan}

\affil[*]{jacek.kasprzak@cnrs.fr}

\begin{abstract}
Coherence transfer is a multi-disciplinary topic of interest, including chemistry, biology and physics. In quantum technologies, achieving non-local coherent coupling between solid-state qubits is of the utmost importance. Here, we demonstrate that excitons — i.e. electron-hole pairs bound by the Coulomb force within a quantum well — can act as a medium for mesoscopic optical coherence transfer in semiconductors. To this end, we use a femtosecond laser pulse to resonantly generate excitons within the light cone. These excitons can then either recombine radiatively or scatter out of the light cone, gaining an in-plane momentum in the process. In samples without disorder, such as the CdTe quantum wells used here, the resulting fast excitons can diffuse over mesoscopic distances before recombining radiatively. Using coherent nonlinear micro-spectroscopy, we carry out exciton time-of-flight measurements. Specifically, we monitor the spatio-temporal propagation of launched exciton wave packets, selectively observing their coherence or density on a scale of up to 10$\,\mu$m. Our proof-of-principle experiment demonstrates that free excitons inherit a phase modulation from the optical pulsed excitation and can generate coherent links within excitonic circuits, offerring a higher level of miniaturisation and compactness than photonic or polaritonic architectures.    
\end{abstract}

\setboolean{displaycopyright}{false} 

\begin{document}
\maketitle
 Controlling the electron transport in semiconductor devices is part of the groundwork of CMOS micro-electronics and computing. Single electron manipulation and transfer along the on-demand circuits can nowadays be achieved\,\cite{HermelinNature11}, enabling quantum optics experiments with electrons\,\cite{WangNatNano23, Ouacel2025}. To realize coherent control in solids on ultrafast, i.e. sub-picosecond time scales, one needs to use optical excitation schemes, usually generating Coulomb bound electron-hole pairs, called excitons, typically hosted by a low-dimensional hetero-structure, like a quantum well (QW). Moving excitons across macroscopic distances is however challenging, because: i) the excitons’ charge neutrality renders them less steerable with external electric fields than electrons, ii) the excitons’ centre of mass (CoM) can be captured on a disorder in a QW\,\cite{SakakiAPL1987, Gaertner2006, SavonaPRB06}, thus hindering the propagation, iii) due to their short lifetime, excitons show an intrinsic non-equilibrium, driven-dissipative nature, requiring optical source.

In practice, low-momentum excitons thus mostly recombine radiatively on pico- to nano-second time scales\,\cite{Deveaud1991}, instead of exhibiting the CoM movement. To overcome the disorder-induced localization of QW excitons, one could replace them with polaritons, the propagation of which has been investigated since early studies on bulk crystals\,\cite{Masumoto1979, Ikehara1991}. Owing to the motional narrowing\,\cite{WhittakerPRL96}, polaritons are less sensitive to microscopic disorder, permitting them to travel across hundreds microns\,\cite{Wertz2010, Song2019, Dang2024}, yet still displaying a lifetime in a pico-second scale owing to its photon part. For the same reason, the polariton coherence volume is more than two orders of magnitude larger than that of the excitons, which could be considered as a disadvantage when it comes to miniaturisation limits. Conversely, working with indirect excitons in double QW systems, the radiative lifetime could be prolonged to micro-seconds\,\cite{Combescot2017}, enabling observation of their diffusion on a scale of one hundred microns\,\cite{Voeroes2005, Fedichkin2016, Nakayama2025}. In a more recent and analogous system, consisting of a hetero bi-layer of transition metal dichalcogenides (TMD), the transport of dark excitons has recently been reported\,\cite{Sun2021}. By reducing the disorder in van der Waals heterostructures and injecting free carriers into the layers, the carrier localization could be suppressed to a level that permits the diffusion of direct (but dark i.e. out-of-the-light-cone) excitons \,\cite{Malic2023, Lee2023} on a hundred microns scale\,\cite{FowlerGerace2021, Aguila2023}. The presence of the electron gas in a QW is known to efficiently screen out the residual disorder, further reducing the spectral inhomogeneous broadening\,\cite{Sanvitto2001, Yamaguchi2008, Rodek2023}, which typically reflects the amount of disorder present in exciton environment.

\begin{figure}[ht!]
    \centering
\includegraphics[width=1.07\columnwidth]{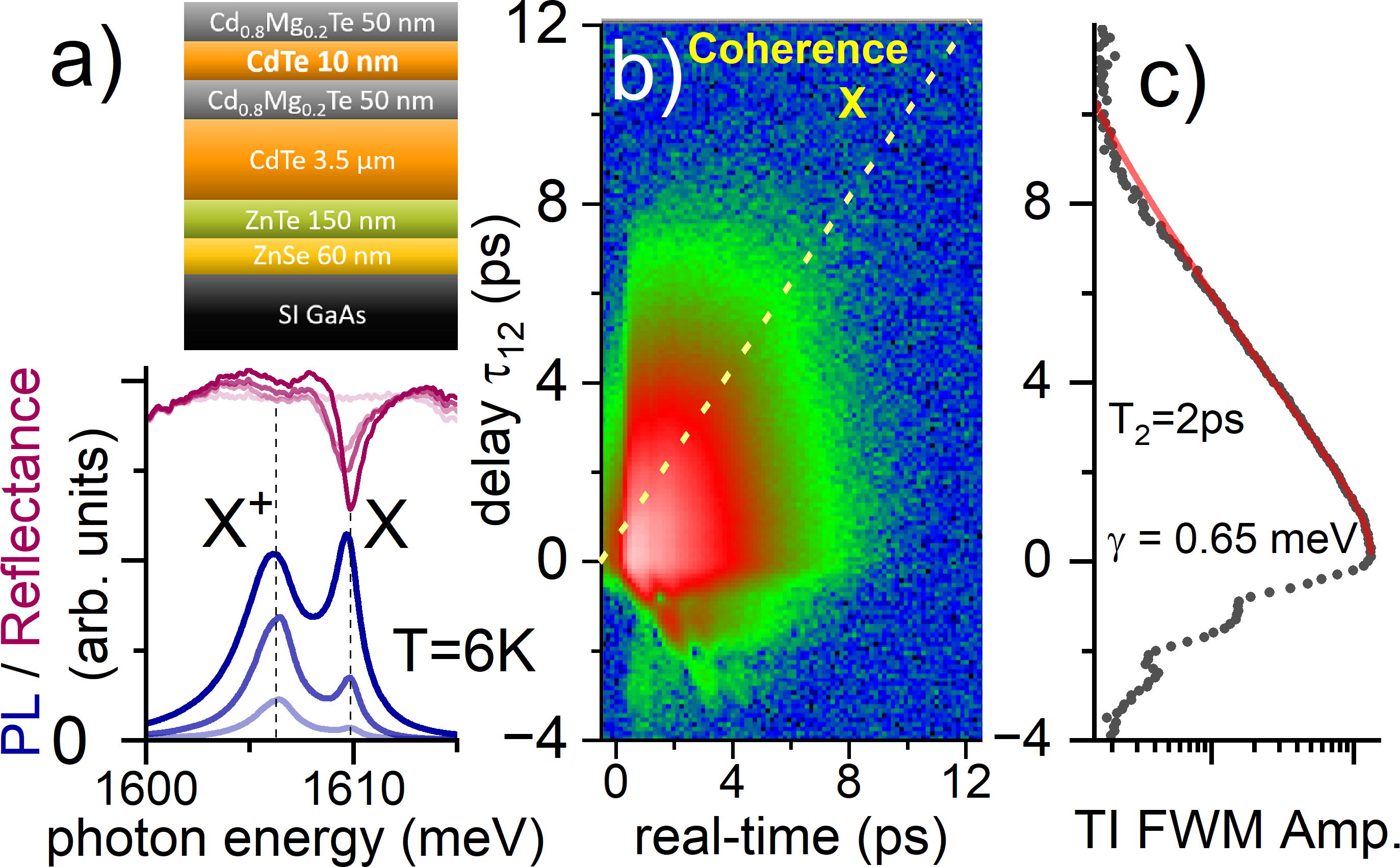}
    \caption{{\bf Linear and nonlinear optical response of the investigated CdTe quantum wells.} a)\,Top:\,Sample layout, Bottom:\,Micro photoluminescence (blue) excited with a 532\,nm CW diode, micro-reflectance (red) measured with a femto-second pulse centered 1610\,meV, revealing the exciton-positive trion complex (X/X+). An increasing 532\,nm excitation intensity encoded from the semi-transparent to solid lines. b)\,Time-resolved FWM of the exciton versus delay $\tau_{12}$, showing the free-induction decay. c)\,Time-integrated (TI) FWM amplitude versus delay $\tau_{12}$ yielding $(T_2,\,\gamma)$=(2\,ps,\,0.65\,meV).}
    \label{fig:1}
\end{figure}

Previous works have mostly explored propagative effects of excitons by measuring the spatio-temporal spread of their non-resonantly excited photoluminescence. In more involved experiments, the exciton\,\cite{Aguila2023}, carrier\,\cite{LafuenteSampietro2018, Ren2022} or magnon\,\cite{Bae2022} diffusion have been probed by using incoherent spatially-resolved pump-probe approaches. Newly, nonlinear spectroscopy was applied to demonstrate non-local coherent coupling between spatially separated polariton states\,\cite{LiuPRL25}. In this Letter, we employ coherent nonlinear spectroscopy to investigate the coherent spatial propagation of QW excitons. To this end, we perform non-local four-wave mixing (FWM) spectroscopy. Heterodyne detection\,\cite{Groll2025} previously permitted us to work in a microscopy configuration, enabling to perform hyperspectral FWM imaging\,\cite{KasprzakNPho11}. Here, by spatially displacing and temporarily delaying respective beams driving the FWM, we perform time-of-flight measurement  of either the exciton density or coherence. 
    
The studied samples consist of a 10\,nm wide CdTe quantum well sandwiched between Cd$_{0.8}$Mg$_{0.2}$Te barriers. Molecular Beam Epitaxy growth of QWs and the sample structure is described in Ref.\,\cite{Bogucki2022}, however here QWs are undoped. We take advantage of the long-standing spectroscopic studies of this material system offering today samples of an excellent optical quality, as witnessed in the photoluminescence and reflectance presented in Fig.\,\ref{fig:1}\,a. The FWM characterization is presented in supplementary Fig.\,S1. Moreover, CdTe QWs host excitons of higher binding energy than the GaAs platform, and thus more robust especially in a presence of free carriers. The positive doping is due to the acceptor levels located on the surface supplying holes penetrating to the QW. The reflectance spectrum obtained at 6\,K with a femtosecond laser pulse is dominated by the exciton-trion complex, similarly as in previous works\,\cite{Ciulin2000}. Thanks to the photo-doping, the absorption ratio between the exciton and the trion could be tuned through the intensity of an auxiliary non-resonant illumination (Fig.\,\ref{fig:1}\,a). We here use a 532\,nm CW laser diode with an average power of only a few $\mu$W, homogeneously covering the sample area of around 50$\mu$m diameter. As such, the density of free carriers of 10$^{11}$\,cm$^{-2}$ is present, efficiently screening the electronic disorder, while performing the FWM experiments. The resident holes in a QW  further screen the residual disorder acting on the excitons, at the same time saturating the non-radiative recombination traps, that both hinder the in-plane propagation of the excitons’ CoM.

\begin{figure*}
    \centering  \includegraphics[width=2.1\columnwidth]{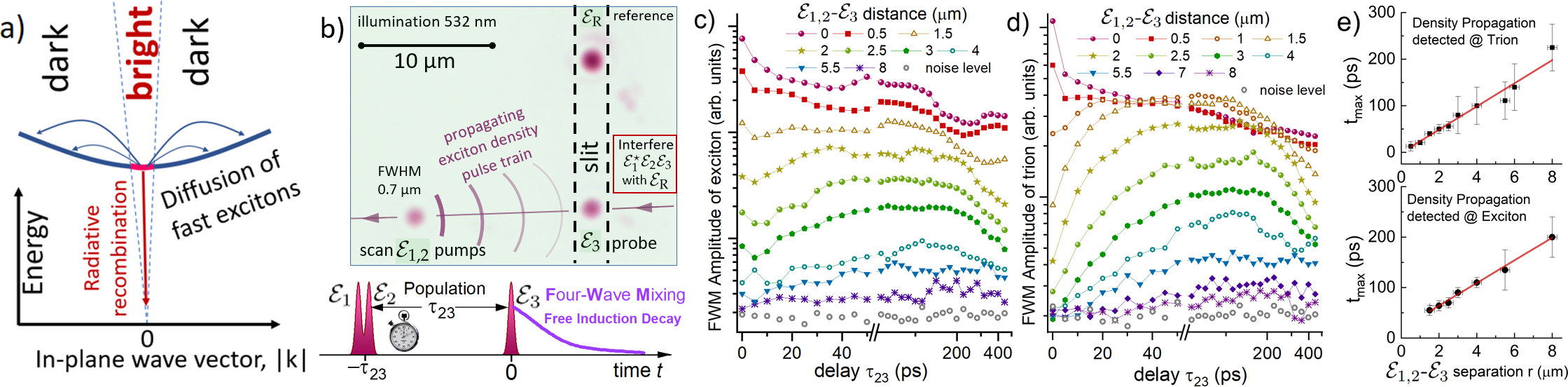}
    \caption{{\bf Density diffusion of QW excitons measured with non-local four-wave mixing microscopy.} a)\,Schematics of the QW excitons' dynamics, depicting an interplay of the direct radiative recombination of bright excitons and scattering towards the dark states, supplying large in-plane momenta and thus enabling a long-range spatial diffusion. b)\,Experimental configuration; Pumps $\Ed$ induce propagation of the exciton's density. The spatio-temporal dynamics of the diffusion process is inferred by measuring FWM amplitude as a function of $\Ed-\Ec$ distance and delay $\tau_{23}$, as illustrated in the pulse sequence. c)\,and d)\,Signatures of the exciton diffusion observed in the FWM ($\tau_{23},\,$r)-dependence, detected at the exciton X and trion X+, respectively. Note that the propagation signature appears to be more pronounced when monitoring X+. This is because X+ absorption, and thus FWM, are highly sensitive to the incoming exciton population. e)\,Time-of-flight analysis yielding the density propagation velocity of $v_{\rm D}=4\times10^4\,$m/s.}
    \label{fig:2}
\end{figure*}

With FWM spectroscopy we can disentangle inhomogeneous and homogeneous broadenings $(\sigma,\,\gamma)$ within the spectral line shape of an optical transition\,\cite{Groll2025}. Under the so-called photon echo configuration, there is a phase conjugation between the first-order polarization and the FWM. If $\sigma>>\gamma$, the signal is a Gaussian centred at $t=\tau_{12}$, marked with a yellow dashed line in Fig.\,\ref{fig:1}\,b, with a time-integrated amplitude decaying in $\tau_{12}$ according to $T_2=2\hbar/\gamma$. When decreasing $\sigma$, the temporal width of the echo increases. When $\sigma<<\gamma$, the temporal rephasing effect becomes negligible and the FWM transient takes the form of an exponential decay, regardless of the delay $\tau_{12}$: this is called the free-induction decay. To check the dominating broadening mechanism, we have spectrally shaped the excitation laser, to selectively excite the heavy-hole exciton, X. The resulting time-resolved FWM is presented in Fig.\,\ref{fig:1}\,b. The data show no traces of the photon echo formation, thus revealing a homogeneously broadened QW system, i.e. $\sigma<<\gamma$. In fact, via raster scanning the FWM transients across the area of 20$\,\mu$m\,x\,10$\,\mu$m, no signs of the photon echo formation could be observed. Free-induction decay dominates, which is a smoking gun of a suppressed disorder on a mesoscopic scale. The time-integrated FWM shown in Fig.\,\ref{fig:1}\,c, displays an exponential decay $\exp(-\tau_{12}/T_2)$, from which we obtain dephasing time of $T_2=2\,$ps and $\gamma=0.65\,$meV (FWHM). Note the presence of the signal for $\tau_{12}<0$ and the quantum beat, both stemming from the exciton-biexciton interaction\,\cite{Langbein2001}, also  indicating the disorder free environment.

As our QW is virtually free from disorder, the excitons experience no spatial localization due to the microscopic potential fluctuations. Under these conditions, the scattering towards the momentum dark-states, illustrated in Fig.\,\ref{fig:2}\,a, occurs via acoustic phonons\,\cite{BorriPRB99B, BorriPRB99} and exciton collisions\,\cite{Kira2006}. The resulting fast excitons can diffuse over mesoscopic distances before relaxing back into the light-cone and recombining radiatively. The excitons with a large in-plane momentum of their CoM (outside the light cone), can propagate in-plane of the quantum well reaching a mesoscopic distance of around 10\,$\mu$m, as reported in Fig.\,\ref{fig:2}\,c,\,d.

We here take advantage of the microscopic configuration of the FWM to probe the diffusion of the exciton density. In contrast to previous FWM studies of exciton diffusion\,\cite{PortellaOberli2002}, we use diffraction limited excitation spots of 0.7\,$\mu$m diamater, permitting us to carry out a time-of-flight experiment directly. To this end, FWM is measured for different spatial separations between the overlapping pumps $\Ed$, and the probe $\Ec$, as depicted in Fig.\,\ref{fig:2}\,b. The two pumps $\Ed$ --- which are 100\,fs laser pulses radio-frequency upshifted by $\Omega_1$ and $\Omega_2$ using an acousto-optic deflection --- impinge the sample, resonantly driving the exciton transition. Subsequent exciton wave-packets, traveling outwards from $\Ed$, generate the temporal density grating modulated at the $\Omega_2-\Omega_1$ frequency. If this grating is sustained during the in-plane propagation, it is converted by the probe beam $\Ec$, modulated at $\Omega_3$, into the FWM at the frequency $\Omega_3+\Omega_2-\Omega_1$. The signal is then  heterodyne detected, with the sensitivity enhanced by spectrally interfering with the reference beam $\Er$, focused 12$\,\mu$m above $\Ec$.

The central result is shown in Figs.\,\ref{fig:2}\,c and d, where the exciton density dynamics, i.e. FWM amplitude versus $\tau_{23}$, is plotted for increasing spatial separation $r$ between $\Ed$ and $\Ec$, and read out at X and X+, respectively. At the spatial overlap of all three beams (the top-most traces) we observe a decrease of the FWM signal on a scale of a few ps, due to a direct radiative decay, followed by secondary recombination of longer lived dark excitons on a nano-second timescale\,\cite{JakubczykACSNano19}, also see Fig.\,S2. With increasing $r$, two main features are observed: Firstly, the signal at $\tau_{23}=0$ is gradually suppressed by more than two orders of magnitude down to the noise level. This indicates a negligible influence of scattered light and Airy fringes originating from the pumps $\Ed$ at the position of the probe beam $\Ec$, as characterized in Fig.\,S3. Secondly, the maximum of the FWM distribution shifts to increasingly longer delays, which is a decisive signature of the diffusive excitons. This experiment is thus an analogue of the time-of-flight-measurement\,\cite{Masumoto1979, McLafferty1981}: a progressively longer lag of the FWM maximum is observed with increasing $r$. Namely, for $r$=4$\,\mu$m the lag-time is around 100\,ps, yielding the exciton group velocity of $v_{\rm d}=40\,$km/s. Which physical processes permit the excitons to reach such $v_{\rm D}$\,? First, the energy of the excitons within the radiative cone, generated at a focus of $0.7\,\mu$m, corresponds to 2.2$\mu$eV, which results in a group velocity of barely 1.5\,km/s. Secondly, at 6\,K, cold excitons generated at the bottom of their dispersion curve get thermally distributed within 0.5\,meV, enabling velocities of up to 12\,km/s to be reached. Thus, to attain $v_{\rm d}$ efficient exciton-acoustic phonon and exciton-exciton scatterings are required, as discussed further. In Fig.\,S4, the same data are presented as a function of $r$ for different delays $\tau_{23}$, intuitively illustrating the spatial diffusion of excitons. In Fig.\,S5, we further verified that the redistribution of the FWM amplitude versus $\tau_{23}$ when varying $r$ originates from the exciton diffusion, and not from the scattered light from $\Ed$ reaching $\Ec$. 

\begin{figure}[ht]
    \centering  \includegraphics[width=1.03\columnwidth]{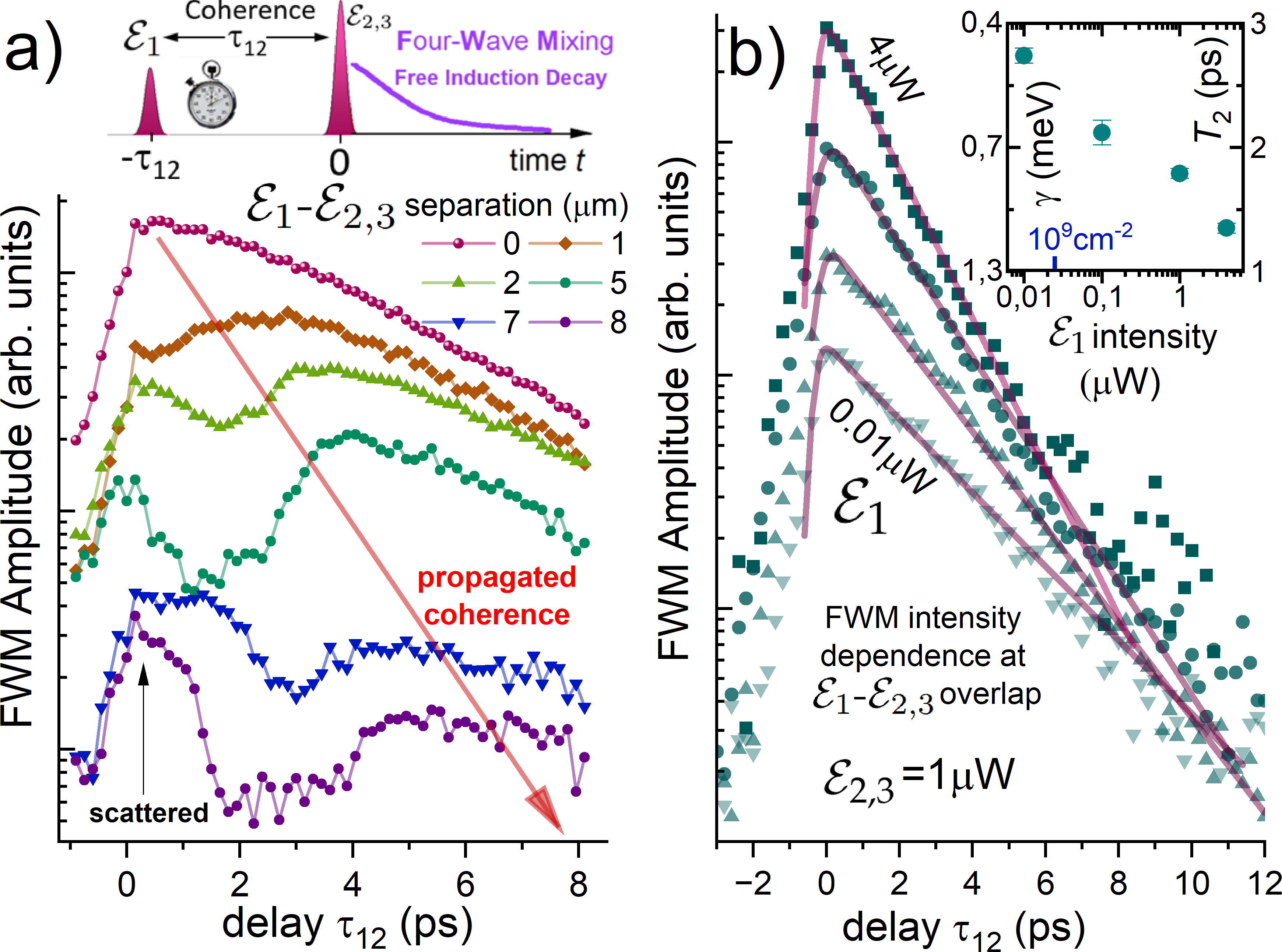}
    \caption{{\bf Coherence propagation of QW excitons (X) measured with non-local four-wave mixing.} a)\,Top: Pulse sequence employed to measure coherence dynamics. Bottom: FWM amplitude of X as a function of $\tau_{12}$. For increasing spatial separation of $\Ea$ and $\Eoo$ a cross-over from an exponential decay toward delayed maximum of the FWM occurs due to coherence propagating from $\Ea$ to $\Eoo$. The leftover FWM signal around $\tau_{12}$=0 is due to direct scattering of $\Ea$ onto $\Eoo$. b)\,$\Ea$ intensity dependence of the coherence dynamics at the beam overlap. A considerable excitation induced dephasing is measured, as quantified in the inset.}
    \label{fig:3}
\end{figure}

To interpret above results, we consider the equation for isotropic diffusion, which in two dimensions and in radial coordinates takes the following form:
\begin{equation}
    \frac{\partial n(r,\,t)}{\partial t}=D\frac{\partial^2 n(r,\,t)}{\partial r^2}+\frac{D}{r}\frac{\partial n(r,\,t)}{\partial r}-\frac{n(r,\,t)}{\tau}
\end{equation}
Here, $D$ is a diffusion coefficient, and $\tau$ is the population radiative and non-radiative decay, see Fig.\,S2 for details. The well-known solution reads:
\begin{equation}
    n(r,\,t)=\frac{n(r=0,t=0)}{4\pi Dt}\exp({\frac{-r^2}{4Dt}}-{\frac{t}{\tau})}
\end{equation} 

The measured FWM amplitude reflects the spatio-temporal dynamics of exciton population, i.e. $n(r,\,t)$. Due to spatially varying carrier concentration and partial beam overlap for small separations between the beams, the resulting FWM could be quite complex.  Nevertheless, the above expression predicts that the exciton distribution will reach a maximum at $t_{\rm max}(r)=\frac{\tau}{2}(\sqrt{1+\frac{r^2}{D\tau}}-1)$. For $r^2\gg D\tau$ this can be approximated with a linear function $t_{\rm max}(r)\approx\frac{\tau}{2}(\frac{r}{\sqrt{D\tau}}-1)$. To corroborate this prediction with our experimental results, we assign the local maxima from the FWM distributions presented in Figs.\,\ref{fig:2}\,c,\,d and plot them in Fig.\,\ref{fig:2}\,e.
The maxima indeed follow a linear dependence with respect to the separation $r$, when monitoring both X and X+. These considerations permit us to estimate the diffusion coefficient: $D\simeq (r/t_{\rm max})^2\tau/4=v_{\rm d}^2\tau/4=16\,$cm$^2$/s, which is well consistent with the previous work on a CdTe QW\,\cite{PortellaOberli2002}. 

The propagation of density measured via FWM micro-spectroscopy presented above is equivalent to the more common incoherent diffusion of excitons, as measured using pump-probe or time-resolved photoluminescence spatial imaging techniques. In the following, we demonstrate the propagation of exciton coherence. FWM spectroscopy is perfectly suited to this task, which however remains challenging owing to dephasing in a ps range. The geometry of the experiment is similar to Fig.\,\ref{fig:2}\,b, while the pulse sequence is depicted in Fig.\,\ref{fig:3}\,a. Here, $\Ea$ generates exciton coherence traveling in the plane of a QW. Upon reaching the overlapping $\Eoo$, the coherence is converted into FWM, interfered with $\Er$ and heterodyne detected. The result of such an experiment for an increasing distance between $\Ea$ and $\Eoo$ is presented in Fig.\,\ref{fig:3}\,a. As the separation between $\Ea$\,-\,$\Eoo$ increases, coherence dynamics diverts from a single exponential decay and develops local maximum which shifts towards positive $\tau_{12}$, to finally form a separate lagged peak. For example, for 5\,$\mu$m separation the measured time-of-flight of exciton coherence is about 4\,ps, from which we deduce coherence propagation velocity of $v_{\rm c}=1250\,$km/s. Such a surprisingly high $v_{\rm c}$, two orders of magnitude higher than $v_{\rm d}$, calls for different mechanisms involved in the exciton density and coherence propagation; The random walk of the exciton density evolving on a nano-second time scale can be modeled by diffusive transport (Eq.\,1). Conversely, coherence is only assigned to bright excitons, meaning it vanishes within the dephasing time of a few picoseconds, indicating that coherence propagation is ballistic, and thereofre $v_{\rm c}\gg v_{\rm d}$. To assess the impact of the scattered light from $\Ea$ onto coherence propagation result, in Fig.\,\ref{fig:3}\,b we present intensity dependence of the coherence dynamics. With increasing $\Ea$ intensity, the coherence decay remains exponential, showing that the FWM redistribution in Fig.\,\ref{fig:3}\,a is indeed due to the coherence propagation. However, the data reveal a considerable excitation-induced dephasing\,\cite{JakubczykACSNano19} (see inset). This confirms the presence of exciton-exciton scattering, which, alongside exciton-phonon scattering, supply excitons with high in-plane momenta.       

To conclude, we took advantage of the strong, coherent nonlinear optical response of a homogeneously broadened CdTe QW to carry out non-local FWM experiments. We demonstrated spatial diffusion of excitons' coherence and density on a length scale of 10\,$\mu$m. Our results demonstrate that FWM micro-spectroscopy can be used to infer the propagative effects of excitons in semiconductor nanostructures. This is particularly impactful in the context of emerging non-local coherent control experiments of individual quantum systems embedded in photonic circuits\,\cite{Tiranov2023}. Our approach could be employed to prove a remote coherent interaction via propagating photons\,\cite{KasprzakNPho11, LiuPRL25}. Another fascinating perspective is to investigate the propagation of coherence and density of valley-excitons in emerging 2D heterostructures with a variable confinement\,\cite{Thureja2022}. Beyond the condensed matter physics, coherent propagation of carriers at the micro- and meso-scale is crucial in biology\,\cite{Rousseau2025}, photovoltaics\,\cite{Bredas2016}, and chemistry\,\cite{DadashiSilab2024}, highlighting the interdisciplinary nature of our work.

\begin{backmatter}
\bmsection{Acknowledgment} We acknowledge the financial support from National Science Centre Poland, projects no. 2023/51/B/ST3/01710 and 2021/41/B/ST3/04183. This work was supported by the NRF of Korea (RS-2023-00236798). We thank Daniel Wigger and Kwangseuk Kyhm for their constructive comments on the manuscritpt.
\bmsection{Disclosures} The authors declare no competing financial interests.
\end{backmatter}
\bibliography{kasprzak19}

@Article{BorriPRB99,
  author   = {P. Borri and W. Langbein and J. M. Hvam and F. Martelli},
  journal  = {Phys. Rev. B},
  title    = {Binding-Energy and Dephasing of Biexcitons in {I}n$_{0.18}${G}a$_{0.82}${A}s/{G}a{A}s Single Quantum Wells},
  year     = {1999},
  pages    = {4505},
  volume   = {60},
}

@Article{BorriPRB99B,
  author   = {P. Borri and W. Langbein and J. M. Hvam and F. Martelli},
  journal  = {Phys. Rev. B},
  title    = {Well-Width Dependence of Exciton-Phonon Scattering in {In$_{x}$Ga$_{1-x}$As/GaAs} Single Quantum Wells},
  year     = {1999},
  pages    = {2215},
  volume   = {59},
}

@Article{JakubczykACSNano19,
  author    = {Tomasz Jakubczyk and Goutham Nayak and Lorenzo Scarpelli and Francesco Masia and Wei-Lai Liu and Sudipta Dubey and Nedjma Bendiab and La\"{e}titia Marty and Takashi Taniguchi and Kenji Watanabe and Gilles Nogues Johann Coraux and Vincent Bouchiat and Wolfgang Langbein and Julien Renard and Jacek Kasprzak},
  title     = {Coherence and density dynamics of excitons in a single-layer {M}o{S}$_2$ reaching the homogeneous limit},
  journal   = {ACS Nano},
  year      = {2019},
  volume    = {13},
  pages     = {3500--3511},
}

@ARTICLE{KasprzakNPho11,
  author = {Kasprzak, Jacek and Patton, Brian and Savona, Vincenzo and Langbein,
	Wolfgang},
  title = {Coherent coupling between distant excitons revealed by two-dimensional
	nonlinear hyperspectral imaging},
  journal = {Nat. Phot.},
  year = {2011},
  volume = {5},
  pages = {57-63}
}

@ARTICLE{SavonaPRB06,
  author = {V.~Savona and W.~Langbein},
  title = {Realistic Heterointerface Model for Excitonic States in Growth-Interrupted
	{GaAs} Quantum Wells},
  journal = {Phys. Rev. B},
  year = {2006},
  volume = {74},
  pages = {75311}
}

@ARTICLE{WhittakerPRL96,
  author = {D. M. Whittaker and P. Kinsler and T. A. Fisher and M. S. Skolnick
	and A. Armitage and A. M. Afshar and M. D. Sturge and J. S. Roberts},
  title = {Motional Narrowing in Semiconductor Microcavities},
  journal = {Phys. Rev. Lett.},
  year = {1996},
  volume = {77},
  pages = {4792-95}
}

@Article{HermelinNature11,
  author    = {Hermelin, Sylvain and Takada, Shintaro and Yamamoto, Michihisa and Tarucha, Seigo and Wieck, Andreas D. and Saminadayar, Laurent and B\"{a}uerle, Christopher and Meunier, Tristan},
  journal   = {Nature},
  title     = {Electrons surfing on a sound wave as a platform for quantum optics with flying electrons},
  year      = {2011},
  issn      = {1476-4687},
  month     = sep,
  number    = {7365},
  pages     = {435--438},
  volume    = {477},
  doi       = {10.1038/nature10416},
  publisher = {Springer Science and Business Media LLC},
}

@Article{WangNatNano23,
  author    = {Wang, Junliang and Edlbauer, Hermann and Richard, Aymeric and Ota, Shunsuke and Park, Wanki and Shim, Jeongmin and Ludwig, Arne and Wieck, Andreas D. and Sim, Heung-Sun and Urdampilleta, Matias and Meunier, Tristan and Kodera, Tetsuo and Kaneko, Nobu-Hisa and Sellier, Hermann and Waintal, Xavier and Takada, Shintaro and B\"{a}uerle, Christopher},
  journal   = {Nature Nanotechnology},
  title     = {Coulomb-mediated antibunching of an electron pair surfing on sound},
  year      = {2023},
  issn      = {1748-3395},
  month     = may,
  number    = {7},
  pages     = {721--726},
  volume    = {18},
  doi       = {10.1038/s41565-023-01368-5},
  publisher = {Springer Science and Business Media LLC},
}

@Article{SakakiAPL1987,
  author    = {Sakaki, H. and Noda, T. and Hirakawa, K. and Tanaka, M. and Matsusue, T.},
  journal   = {Applied Physics Letters},
  title     = {Interface roughness scattering in GaAs/AlAs quantum wells},
  year      = {1987},
  issn      = {1077-3118},
  month     = dec,
  number    = {23},
  pages     = {1934--1936},
  volume    = {51},
  doi       = {10.1063/1.98305},
  publisher = {AIP Publishing},
}

@Article{Gaertner2006,
  author    = {G\"{a}rtner, A. and Holleitner, A. W. and Kotthaus, J. P. and Schuh, D.},
  journal   = {Applied Physics Letters},
  title     = {Drift mobility of long-living excitons in coupled {G}a{A}s quantum wells},
  year      = {2006},
  issn      = {1077-3118},
  month     = jul,
  number    = {5},
  volume    = {89},
  doi       = {10.1063/1.2267263},
  publisher = {AIP Publishing},
}

@Article{Wertz2010,
  author    = {Wertz, E. and Ferrier, L. and Solnyshkov, D. D. and Johne, R. and Sanvitto, D. and Lemaître, A. and Sagnes, I. and Grousson, R. and Kavokin, A. V. and Senellart, P. and Malpuech, G. and Bloch, J.},
  journal   = {Nature Physics},
  title     = {Spontaneous formation and optical manipulation of extended polariton condensates},
  year      = {2010},
  issn      = {1745-2481},
  month     = aug,
  number    = {11},
  pages     = {860--864},
  volume    = {6},
  doi       = {10.1038/nphys1750},
  publisher = {Springer Science and Business Media LLC},
}

@Article{Combescot2017,
  author    = {Combescot, Monique and Combescot, Roland and Dubin, François},
  journal   = {Reports on Progress in Physics},
  title     = {BoseEinstein condensation and indirect excitons: a review},
  year      = {2017},
  issn      = {1361-6633},
  month     = mar,
  number    = {6},
  pages     = {066501},
  volume    = {80},
  doi       = {10.1088/1361-6633/aa50e3},
  publisher = {IOP Publishing},
}

@Article{Voeroes2005,
  author    = {V\"{o}r\"{o}s, Z. and Balili, R. and Snoke, D. W. and Pfeiffer, L. and West, K.},
  journal   = {Physical Review Letters},
  title     = {Long-Distance Diffusion of Excitons in Double Quantum Well Structures},
  year      = {2005},
  issn      = {1079-7114},
  month     = jun,
  number    = {22},
  pages     = {226401},
  volume    = {94},
  doi       = {10.1103/physrevlett.94.226401},
  publisher = {American Physical Society (APS)},
}

@Article{Sun2021,
  author    = {Sun, Zhe and Ciarrocchi, Alberto and Tagarelli, Fedele and Gonzalez Marin, Juan Francisco and Watanabe, Kenji and Taniguchi, Takashi and Kis, Andras},
  journal   = {Nature Photonics},
  title     = {Excitonic transport driven by repulsive dipolar interaction in a van der {W}aals heterostructure},
  year      = {2021},
  issn      = {1749-4893},
  month     = dec,
  number    = {1},
  pages     = {79--85},
  volume    = {16},
  doi       = {10.1038/s41566-021-00908-6},
  publisher = {Springer Science and Business Media LLC},
}

@Article{Malic2023,
  author    = {Malic, Ermin and Perea-Causin, Raül and Rosati, Roberto and Erkensten, Daniel and Brem, Samuel},
  journal   = {Nature Communications},
  title     = {Exciton transport in atomically thin semiconductors},
  year      = {2023},
  issn      = {2041-1723},
  month     = jun,
  number    = {1},
  volume    = {14},
  doi       = {10.1038/s41467-023-38556-9},
  publisher = {Springer Science and Business Media LLC},
}

@Article{Lee2023,
  author    = {Lee, Hyeongwoo and Kim, Yong Bin and Ryu, Jae Won and Kim, Sujeong and Bae, Jinhyuk and Koo, Yeonjeong and Jang, Donghoon and Park, Kyoung-Duck},
  journal   = {Nano Convergence},
  title     = {Recent progress of exciton transport in two-dimensional semiconductors},
  year      = {2023},
  issn      = {2196-5404},
  month     = dec,
  number    = {1},
  volume    = {10},
  doi       = {10.1186/s40580-023-00404-3},
  publisher = {Springer Science and Business Media LLC},
}

@Article{Aguila2023,
  author    = {del Aguila, Andr\'{e}s Granados and Wong, Yi Ren and Wadgaonkar, Indrajit and Fieramosca, Antonio and Liu, Xue and Vaklinova, Kristina and Dal Forno, Stefano and Do, T. Thu Ha and Wei, Ho Yi and Watanabe, K. and Taniguchi, T. and Novoselov, Kostya S. and Koperski, Maciej and Battiato, Marco and Xiong, Qihua},
  journal   = {Nature Nanotechnology},
  title     = {Ultrafast exciton fluid flow in an atomically thin {M}o{S}$_2$ semiconductor},
  year      = {2023},
  issn      = {1748-3395},
  month     = jul,
  number    = {9},
  pages     = {1012--1019},
  volume    = {18},
  doi       = {10.1038/s41565-023-01438-8},
  publisher = {Springer Science and Business Media LLC},
}

@Article{Sanvitto2001,
  author  = {Sanvitto, Daniele and Pulizzi, Fabio and Shields, Andrew J. and Christianen, Peter C. M. and Holmes, Stuart N. and Simmons, Michelle Y. and Ritchie, David A. and Maan, Jan C. and Pepper, Michael},
  journal = {Science},
  title   = {Observation of Charge Transport by Negatively Charged Excitons},
  year    = {2001},
  issn    = {1095-9203},
  number  = {5543},
  pages   = {837--839},
  volume  = {294},
}

@Article{FowlerGerace2021,
  author    = {Fowler-Gerace, L. H. and Choksy, D. J. and Butov, L. V.},
  journal   = {Physical Review B},
  title     = {Voltage-controlled long-range propagation of indirect excitons in a van der {W}aals heterostructure},
  year      = {2021},
  issn      = {2469-9969},
  month     = oct,
  number    = {16},
  pages     = {165302},
  volume    = {104},
  doi       = {10.1103/physrevb.104.165302},
  publisher = {American Physical Society (APS)},
}

@Article{Yamaguchi2008,
  author    = {Yamaguchi, M. and Nomura, S. and Maruyama, T. and Miyashita, S. and Hirayama, Y. and Tamura, H. and Akazaki, T.},
  journal   = {Physical Review Letters},
  title     = {Evidence of a Transition from Nonlinear to Linear Screening of a Two-Dimensional Electron System Detected by Photoluminescence Spectroscopy},
  year      = {2008},
  issn      = {1079-7114},
  month     = nov,
  number    = {20},
  pages     = {207401},
  volume    = {101},
  doi       = {10.1103/physrevlett.101.207401},
  publisher = {American Physical Society (APS)},
}

@Article{Rodek2023,
  author    = {Rodek, Aleksander and Hahn, Thilo and Howarth, James and Taniguchi, Takashi and Watanabe, Kenji and Potemski, Marek and Kossacki, Piotr and Wigger, Daniel and Kasprzak, Jacek},
  journal   = {2D Materials},
  title     = {Controlled coherent-coupling and dynamics of exciton complexes in a {M}o{S}e$_2$ monolayer},
  year      = {2023},
  issn      = {2053-1583},
  month     = apr,
  number    = {2},
  pages     = {025027},
  volume    = {10},
  doi       = {10.1088/2053-1583/acc59a},
  publisher = {IOP Publishing},
}

@Article{LafuenteSampietro2018,
  author    = {Lafuente-Sampietro, A. and Utsumi, H. and Sunaga, M. and Makita, K. and Boukari, H. and Kuroda, S. and Besombes, L.},
  journal   = {Physical Review B},
  title     = {Dynamics of a {C}r spin in a semiconductor quantum dot: Hole-{C}r flip-flops and spin-phonon coupling},
  year      = {2018},
  issn      = {2469-9969},
  month     = apr,
  number    = {15},
  pages     = {155301},
  volume    = {97},
  doi       = {10.1103/physrevb.97.155301},
  publisher = {American Physical Society (APS)},
}

@Article{Ren2022,
  author    = {Ren, L. and Lombez, L. and Robert, C. and Beret, D. and Lagarde, D. and Urbaszek, B. and Renucci, P. and Taniguchi, T. and Watanabe, K. and Crooker, S.?A. and Marie, X.},
  journal   = {Physical Review Letters},
  title     = {Optical Detection of Long Electron Spin Transport Lengths in a Monolayer Semiconductor},
  year      = {2022},
  issn      = {1079-7114},
  month     = jul,
  number    = {2},
  pages     = {027402},
  volume    = {129},
  doi       = {10.1103/physrevlett.129.027402},
  publisher = {American Physical Society (APS)},
}

@Article{Bae2022,
  author    = {Bae, Youn Jue and Wang, Jue and Scheie, Allen and Xu, Junwen and Chica, Daniel G. and Diederich, Geoffrey M. and Cenker, John and Ziebel, Michael E. and Bai, Yusong and Ren, Haowen and Dean, Cory R. and Delor, Milan and Xu, Xiaodong and Roy, Xavier and Kent, Andrew D. and Zhu, Xiaoyang},
  journal   = {Nature},
  title     = {Exciton-coupled coherent magnons in a 2{D} semiconductor},
  year      = {2022},
  issn      = {1476-4687},
  month     = sep,
  number    = {7926},
  pages     = {282--286},
  volume    = {609},
  doi       = {10.1038/s41586-022-05024-1},
  publisher = {Springer Science and Business Media LLC},
}

@Article{Tiranov2023,
  author    = {Tiranov, Alexey and Angelopoulou, Vasiliki and van Diepen, Cornelis Jacobus and Schrinski, Björn and Sandberg, Oliver August DallAlba and Wang, Ying and Midolo, Leonardo and Scholz, Sven and Wieck, Andreas Dirk and Ludwig, Arne and S{\o}rensen, Anders S?ndberg and Lodahl, Peter},
  journal   = {Science},
  title     = {Collective super- and subradiant dynamics between distant optical quantum emitters},
  year      = {2023},
  issn      = {1095-9203},
  month     = jan,
  number    = {6630},
  pages     = {389--393},
  volume    = {379},
  doi       = {10.1126/science.ade9324},
  publisher = {American Association for the Advancement of Science (AAAS)},
}

@Article{Bogucki2022,
  author    = {Bogucki, A. and Goryca, M. and {\L}opion, A. and Pacuski, W. and Po{\l}czy\'{n}ska, K. E. and Domaga{\l}a, J. Z. and Tokarczyk, M. and F\c{a}s, T. and Golnik, A. and Kossacki, P.},
  journal   = {Physical Review B},
  title     = {Angle-resolved optically detected magnetic resonance as a tool for strain determination in nanostructures},
  year      = {2022},
  issn      = {2469-9969},
  month     = feb,
  number    = {7},
  pages     = {075412},
  volume    = {105},
  doi       = {10.1103/physrevb.105.075412},
  publisher = {American Physical Society (APS)},
}

@Article{McLafferty1981,
  author    = {McLafferty, Fred W.},
  journal   = {Science},
  title     = {Tandem Mass Spectrometry},
  year      = {1981},
  issn      = {1095-9203},
  month     = oct,
  number    = {4518},
  pages     = {280--287},
  volume    = {214},
  doi       = {10.1126/science.7280693},
  publisher = {American Association for the Advancement of Science (AAAS)},
}

@Article{Deveaud1991,
  author    = {Deveaud, B. and Clérot, F. and Roy, N. and Satzke, K. and Sermage, B. and Katzer, D.},
  journal   = {Physical Review Letters},
  title     = {Enhanced radiative recombination of free excitons in {G}a{A}s quantum wells},
  year      = {1991},
  issn      = {0031-9007},
  month     = oct,
  number    = {17},
  pages     = {2355--2358},
  volume    = {67},
  doi       = {10.1103/physrevlett.67.2355},
  publisher = {American Physical Society (APS)},
}

@Article{Fedichkin2016,
  author    = {Fedichkin, F. and Guillet, T. and Valvin, P. and Jouault, B. and Brimont, C. and Bretagnon, T. and Lahourcade, L. and Grandjean, N. and Lefebvre, P. and Vladimirova, M.},
  journal   = {Phys. Rev. Appl.},
  title     = {Room-Temperature Transport of Indirect Excitons in $(\mathrm{Al},\mathrm{Ga})\mathrm{N}/\mathrm{GaN}$ Quantum Wells},
  year      = {2016},
  month     = {Jul},
  pages     = {014011},
  volume    = {6},
  doi       = {10.1103/PhysRevApplied.6.014011},
  issue     = {1},
  numpages  = {10},
  publisher = {American Physical Society},
  url       = {https://link.aps.org/doi/10.1103/PhysRevApplied.6.014011},
}

@Article{Dang2024,
  author    = {Dang, Nguyen Ha My and Zanotti, Simone and Drouard, Emmanuel and Chevalier, Céline and Trippé-Allard, Gaëlle and Deleporte, Emmanuelle and Seassal, Christian and Gerace, Dario and Nguyen, Hai Son},
  journal   = {Nano Letters},
  title     = {Long-Range Ballistic Propagation of 80\% Excitonic Fraction Polaritons in a Perovskite Metasurface at Room Temperature},
  year      = {2024},
  issn      = {1530-6992},
  month     = sep,
  number    = {38},
  pages     = {11839--11846},
  volume    = {24},
  doi       = {10.1021/acs.nanolett.4c02696},
  publisher = {American Chemical Society (ACS)},
}

@Article{Thureja2022,
  author    = {Thureja, Deepankur and Imamoglu, Atac and Smolenski, Tomasz and Amelio, Ivan and Popert, Alexander and Chervy, Thibault and Lu, Xiaobo and Liu, Song and Barmak, Katayun and Watanabe, Kenji and Taniguchi, Takashi and Norris, David J. and Kroner, Martin and Murthy, Puneet A.},
  journal   = {Nature},
  title     = {Electrically tunable quantum confinement of neutral excitons},
  year      = {2022},
  issn      = {1476-4687},
  month     = may,
  number    = {7913},
  pages     = {298--304},
  volume    = {606},
  doi       = {10.1038/s41586-022-04634-z},
  publisher = {Springer Science and Business Media LLC},
}

@Article{Rousseau2025,
  author    = {Rousseau, Adrien and Richardson, Katherine H. and Nandy, Atanu and Vasilev, Cvetelin and Hoffmann, Madeline P. and Hunter, C. Neil and Johnson, Matthew P. and Schlau-Cohen, Gabriela S.},
  journal   = {ACS Nano},
  title     = {Exciton-Diffusion Enhanced Energy Capture in an Integrated Nanoscale Platform},
  year      = {2025},
  issn      = {1936-086X},
  month     = apr,
  number    = {15},
  pages     = {14865--14872},
  volume    = {19},
  doi       = {10.1021/acsnano.4c18713},
  publisher = {American Chemical Society (ACS)},
}

@Article{Bredas2016,
  author    = {Brédas, Jean-Luc and Sargent, Edward H. and Scholes, Gregory D.},
  journal   = {Nature Materials},
  title     = {Photovoltaic concepts inspired by coherence effects in photosynthetic systems},
  year      = {2016},
  issn      = {1476-4660},
  month     = dec,
  number    = {1},
  pages     = {35--44},
  volume    = {16},
  doi       = {10.1038/nmat4767},
  publisher = {Springer Science and Business Media LLC},
}

@Article{DadashiSilab2024,
  author    = {Dadashi-Silab, Sajjad and Preston-Herrera, Cristina and Oblinsky, Daniel G. and Scholes, Gregory D. and Stache, Erin E.},
  journal   = {Journal of the American Chemical Society},
  title     = {Red-Light-Induced Ligand-to-Metal Charge Transfer Catalysis by Tuning the Axial Coordination of Cobyrinate},
  year      = {2024},
  issn      = {1520-5126},
  month     = dec,
  number    = {50},
  pages     = {34583--34590},
  volume    = {146},
  doi       = {10.1021/jacs.4c12432},
  publisher = {American Chemical Society (ACS)},
}

@Article{Langbein2001,
  author    = {Langbein, W. and Meier, T. and Koch, S. W. and Hvam, J. M.},
  journal   = {Journal of the Optical Society of America B},
  title     = {Spectral signatures of $\chi^{(5)}$ processes in four-wave mixing of homogeneously broadened excitons},
  year      = {2001},
  issn      = {1520-8540},
  month     = sep,
  number    = {9},
  pages     = {1318},
  volume    = {18},
  doi       = {10.1364/josab.18.001318},
  publisher = {Optica Publishing Group},
}

@Article{Ikehara1991,
  author    = {Ikehara, T. and Itoh, T.},
  journal   = {Physical Review B},
  title     = {Dynamical behavior of the exciton polariton in {C}u{C}l: Coherent propagation and momentum relaxation},
  year      = {1991},
  issn      = {1095-3795},
  month     = nov,
  number    = {17},
  pages     = {9283--9294},
  volume    = {44},
  doi       = {10.1103/physrevb.44.9283},
  publisher = {American Physical Society (APS)},
}

@Article{Ouacel2025,
  author    = {Ouacel, Seddik and Mazzella, Lucas and Kloss, Thomas and Aluffi, Matteo and Vasselon, Thomas and Edlbauer, Hermann and Wang, Junliang and Geffroy, Clément and Shaju, Jashwanth and Ludwig, Arne and Wieck, Andreas D. and Yamamoto, Michihisa and Pomaranski, David and Takada, Shintaro and Kaneko, Nobu-Hisa and Georgiou, Giorgos and Waintal, Xavier and Urdampilleta, Matias and Sellier, Hermann and Bäuerle, Christopher},
  journal   = {Nature Communications},
  title     = {Electronic interferometry with ultrashort plasmonic pulses},
  year      = {2025},
  issn      = {2041-1723},
  month     = may,
  number    = {1},
  volume    = {16},
  doi       = {10.1038/s41467-025-58939-4},
  publisher = {Springer Science and Business Media LLC},
}

@Article{Masumoto1979,
  author    = {Masumoto, Yasuaki and Unuma, Yutaka and Tanaka, Yuichi and Shionoya, Shigeo},
  journal   = {Journal of the Physical Society of Japan},
  title     = {Picosecond Time of Flight Measurements of Excitonic Polariton in {C}u{C}l},
  year      = {1979},
  issn      = {1347-4073},
  month     = dec,
  number    = {6},
  pages     = {1844--1849},
  volume    = {47},
  doi       = {10.1143/jpsj.47.1844},
  publisher = {Physical Society of Japan},
}

@Article{Song2019,
  author    = {Song, Hyun Gyu and Choi, Sunghan and Park, Chung Hyun and Gong, Su-Hyun and Lee, Chulwon and Kwon, Min Sik and Choi, Dae Gwang and Woo, Kie Young and Cho, Yong-Hoon},
  journal   = {Optica},
  title     = {Tailoring the potential landscape of room-temperature single-mode whispering gallery polariton condensate},
  year      = {2019},
  issn      = {2334-2536},
  month     = oct,
  number    = {10},
  pages     = {1313},
  volume    = {6},
  doi       = {10.1364/optica.6.001313},
  publisher = {Optica Publishing Group},
}

@Article{Nakayama2025,
  author    = {Nakayama, Masaaki and Miyazaki, Yuichiro},
  journal   = {Journal of the Physical Society of Japan},
  title     = {Experimental Verification of Ballistic Propagation of Polaritons in ExcitonExciton Inelastic Scattering in a GaAs/AlAs Multiple-Quantum-Well Structure},
  year      = {2025},
  issn      = {1347-4073},
  month     = jun,
  number    = {6},
  volume    = {94},
  doi       = {10.7566/jpsj.94.064703},
  publisher = {Physical Society of Japan},
}

@Article{LiuPRL25,
  author    = {Liu, Albert and Martin, Eric W. and Hu, Jiaqi and Wang, Zhaorong and Deng, Hui and Cundiff, Steven T.},
  journal   = {Physical Review Letters},
  title     = {Nonlocal Coherent Optical Nonlinearities of a Macroscopic Quantum System},
  year      = {2025},
  issn      = {1079-7114},
  month     = oct,
  number    = {18},
  pages     = {183801},
  volume    = {135},
  doi       = {10.1103/jrds-3tyk},
  publisher = {American Physical Society (APS)},
}

@Article{Groll2025,
  author    = {Groll, Daniel and Hahn, Thilo and Machnikowski, Pawe³ and Kuhn, Tilmann and Kasprzak, Jacek and Wigger, Daniel},
  journal   = {Nano Futures},
  title     = {Fundamentals of heterodyne wave mixing spectroscopy: a tutorial},
  year      = {2025},
  issn      = {2399-1984},
  month     = nov,
  number    = {4},
  pages     = {042601},
  volume    = {9},
  doi       = {10.1088/2399-1984/ae108f},
  publisher = {IOP Publishing},
}

@Article{Ciulin2000,
  author    = {Ciulin, V. and Kossacki, P. and Haacke, S. and Gani?re, J.-D. and Deveaud, B. and Esser, A. and Kutrowski, M. and Wojtowicz, T.},
  journal   = {Physical Review B},
  title     = {Radiative behavior of negatively charged excitons in CdTe-based quantum wells: A spectral and temporal analysis},
  year      = {2000},
  issn      = {1095-3795},
  month     = dec,
  number    = {24},
  pages     = {R16310--R16313},
  volume    = {62},
  doi       = {10.1103/physrevb.62.r16310},
  publisher = {American Physical Society (APS)},
}

@Article{Kira2006,
  author    = {Kira, M. and Koch, S.W.},
  journal   = {Progress in Quantum Electronics},
  title     = {Many-body correlations and excitonic effects in semiconductor spectroscopy},
  year      = {2006},
  issn      = {0079-6727},
  number    = {5},
  pages     = {155--296},
  volume    = {30},
  doi       = {10.1016/j.pquantelec.2006.12.002},
  publisher = {Elsevier BV},
}

@Article{PortellaOberli2002,
  author    = {Portella-Oberli, M. T. and Ciulin, V. and Haacke, S. and Gani?re, J.-D. and Kossacki, P. and Kutrowski, M. and Wojtowicz, T. and Deveaud, B.},
  journal   = {Physical Review B},
  title     = {Diffusion, localization, and dephasing of trions and excitons in CdTe quantum wells},
  year      = {2002},
  issn      = {1095-3795},
  month     = oct,
  number    = {15},
  pages     = {155305},
  volume    = {66},
  doi       = {10.1103/physrevb.66.155305},
  publisher = {American Physical Society (APS)},
}

\subsection{\Large Supplementary figures}

\renewcommand{\thefigure}{S\arabic{figure}}
\setcounter{figure}{0}

\begin{figure}[htbp]
\centering
\fbox{\includegraphics[width=1.05\linewidth]{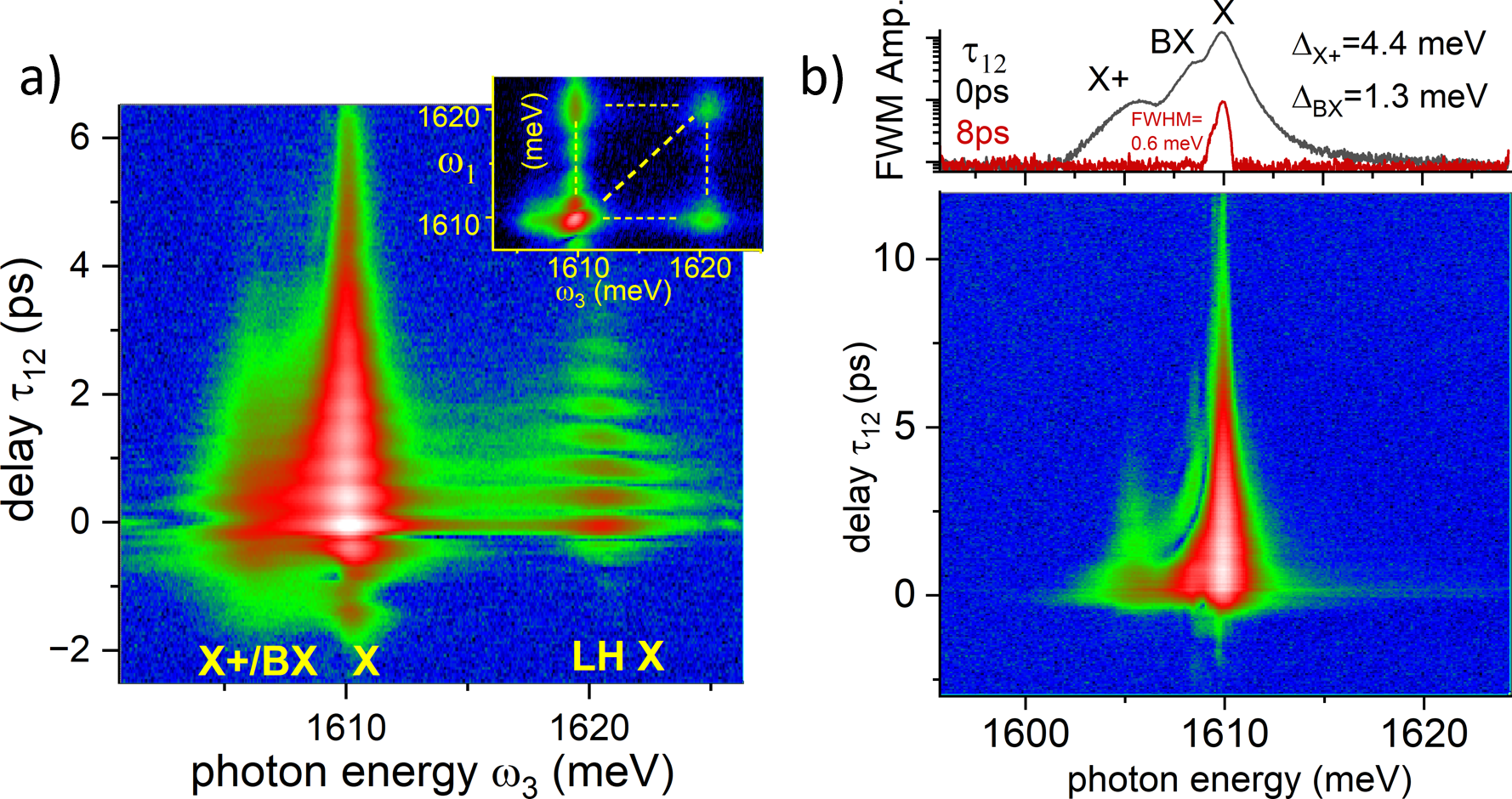}}
\caption{{\bf Four-wave mixing characterisation of the studied CdTe quantum well.} a)\,Four-wave mixing amplitude spectra for different delays $\tau_{12}$ under spectrally broad excitation (25\,meV pulse width centered at 1615\,meV), showing neutral exciton (X), positive trion (X+) and biexciton (BX), and a light-hole neutral exciton (LH X)  transitions. The pronounced quantum beating along the $\tau_{12}$ axis is due to the coherent coupling between the heavy-hole and light-hole exciton transitions. Fourier transforming along $\tau_{12}$ yields a two-dimensional FWM spectrum (see inset) with off-diagonal peaks that confirm the coherent coupling. The horizontally offset signal from X towards the low-energy side is due to the BX. b)\,As a) but under spectrally narrow excitation conditions (10\,meV pulse width centered at 1610\,meV; LH X is not excited). The FWM spectrum at $\tau_{12}=0\,$ps (black) directly shows the X/BX/X+ exciton complex. The FWM spectrum for longer delays ($\tau_{12}=8\,$ps, red) displays a spectral width of 0.6\,meV (FWHM), which corresponds to the homogeneous line width measured via $\tau_{12}$-dependence of $T_2$=2.05\,ps. The logarithmic colour scale covers 2 orders of magnitude from blue to white.}
\label{fig:S1}
\end{figure}

\begin{figure}[htbp]
\centering
\fbox{\includegraphics[width=1.05\linewidth]{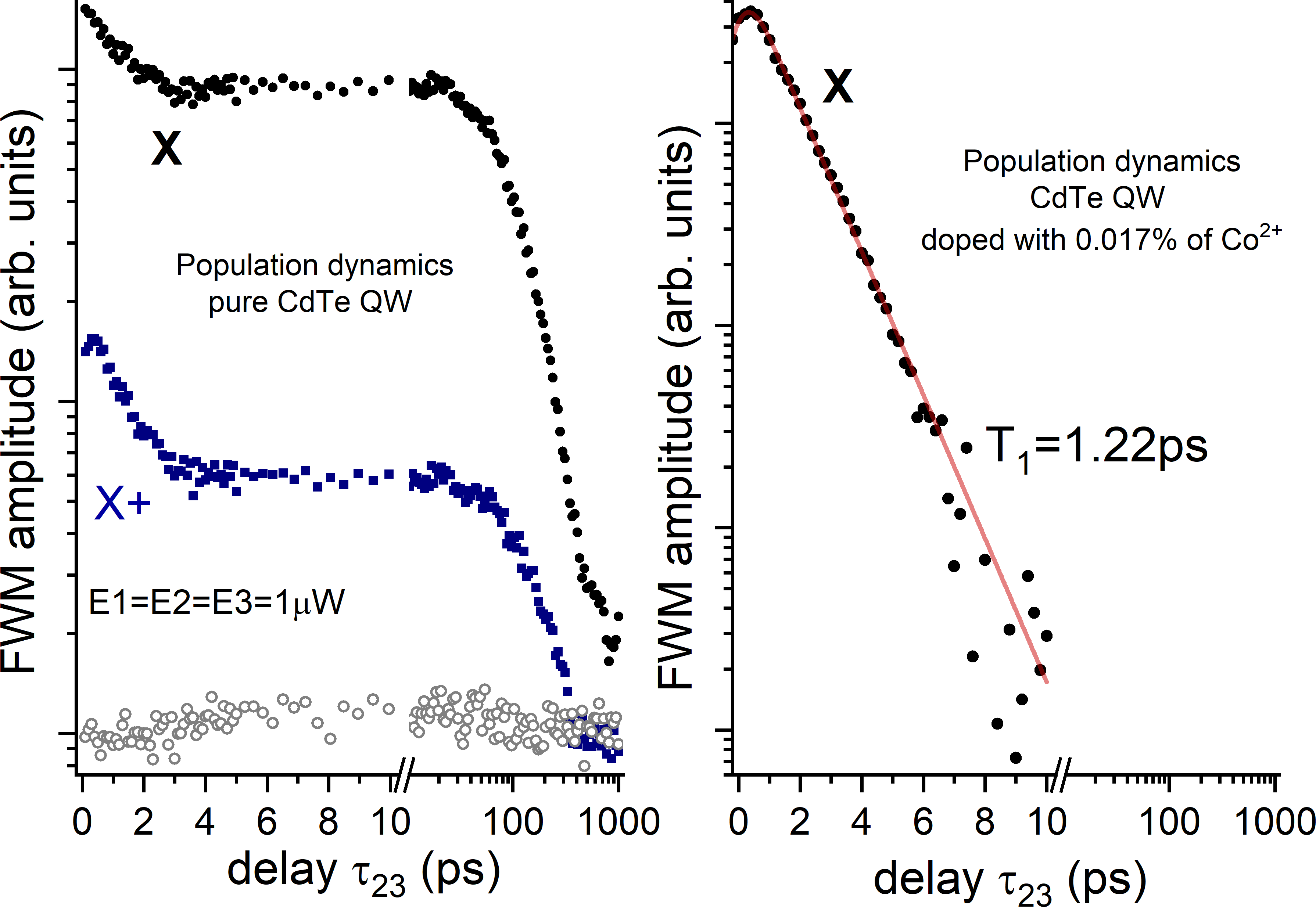}}
\caption{{\bf Population dynamics measured at the beam overlap.} a)\,FWM amplitude versus $\tau_{23}$ on X and X+ transitions measured at a pure CdTe quantum well as employed in the main manuscript. The initial decay on a 4\,ps scale is followed by a re-population of bright exciton states occurring on a time scale of a few hundred ps. b)\,As in a) but measured on a CdTe quantum well doped with 0.17\,\% of Co$^{2+}$ ions. The cobalt ions act as efficient centres for non-radiative exciton recombination, preventing the formation of a long-lived exciton population. This allows us to measure the population lifetime of the initially injected cold excitons, $T_1$=1.22\,ps, approaching the radiative limit $T_1$=$T_2$/2.}
\label{fig:S2}
\end{figure}

\begin{figure}[htbp]
\centering
\fbox{\includegraphics[width=1.03\linewidth]{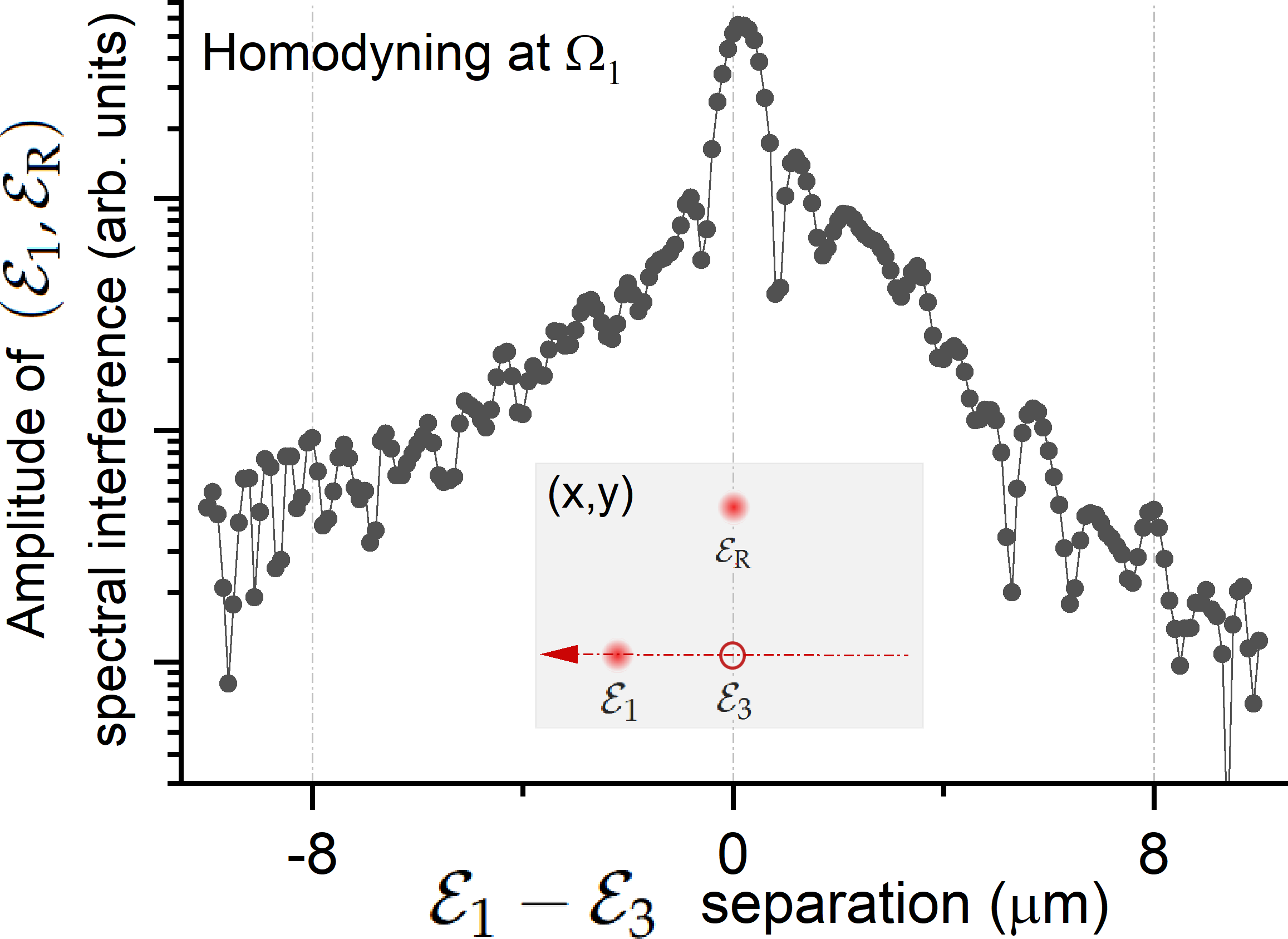}}
\caption{{\bf Characterization of the pump-probe ($\Ea-\Ec$) cross-talk in a time-of-flight experiment.} The amplitude of the pump-reference ($\Ea-\Er$) spectral interference versus pump-probe separation. The geometry of the experiment is shown in the inset. The first Airy ring is clearly visible (one order of magnitude below the maximum of the Airy disk) when approaching the beam overlap. The scattered light from the pump $\Ea$ arrives at the position of the probe $\Ec$. It is then picked up by the reference $\Er$ and generates a spectral interference when homodyning at the $\Ea$ pump frequency $\Omega_1$. Note the logarithmic vertical scale. This characterisation reflects the beam cross-talk in the coherence propagation experiment, in which only one pump $\Ea$ is displaced from the probe $\Eoo$. Indeed, in Fig.\,4\,a some parasitic signal remains visible at zero delay even after displacing the pump by 8\,$\mu$m. In the density propagation experiment reported in Fig.\,2, two pumps $\Ed$ are displaced from the probe $\Ec$, the suppression is therefore accordingly stronger, as the amplitude reported in the above characterisation should be squared.}
\label{fig:S3}
\end{figure}

\begin{figure}[htbp]
\centering
\fbox{\includegraphics[width=1.05\linewidth]{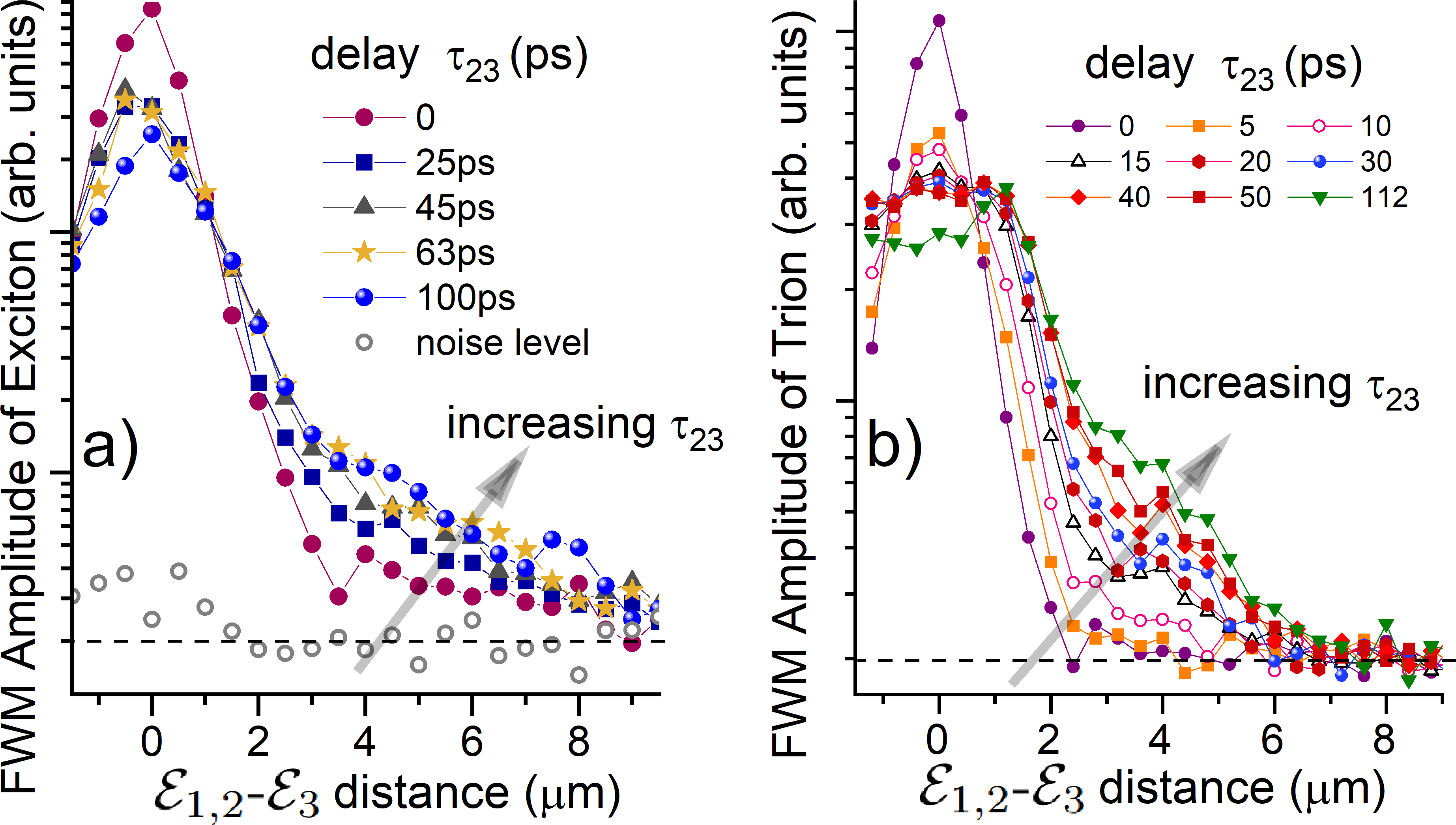}}
\caption{{\bf Spatial diffusion detected on X (left) and X+ (right).} The same data as in Figure 2 c) and d), but ploted as a function of the pumps-probe distance. With increasing $\tau_{23}$ the width of the spatial profile of the FWM increases, intuitively illustrating the spatial diffusion process.}
\label{fig:S4}
\end{figure}

\begin{figure}[htbp]
\centering
\fbox{\includegraphics[width=1.02\linewidth]{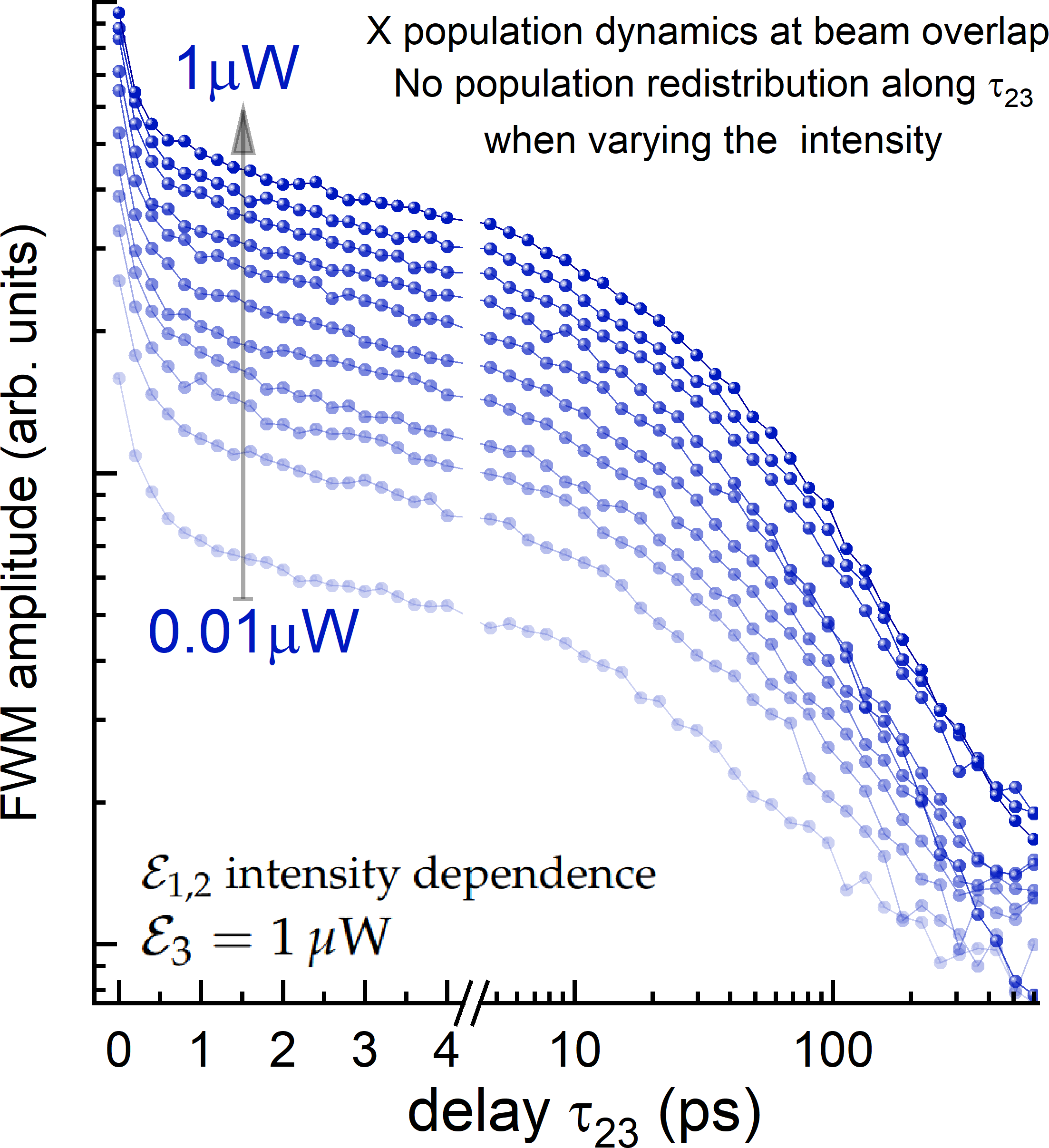}}
\caption{{\bf Population dynamics measured at the beam overlap. Intensity dependence.} The exciton population dynamics measured for a fixed probe intensity $\Ec$ of $1\,\mu$W while varying the intensity of the pumps $\Ed$. Apart from the overall scaling factor, no signs of FWM redistribution were observed when the pump intensities were varied from $0.01\,\mu$W to 1$\,\mu$W. This indicates that exciton transport is the primary effect occurring in the non-local four-wave mixing experiment.}
\label{fig:S5}
\end{figure}

\end{document}